\documentstyle[12pt,epsfig]{article}
\newlength{\dinwidth}
\newlength{\dinmargin}
\begin{document}
\pagestyle{empty}
\thispagestyle{empty}
\vspace{1 cm}
\def\alphas{$\alpha_s$}
\newcommand{\ptrat}     {\mbox{$\frac{p_{Th}}{p_{Te}}$}}
\newcommand{\yjbrat}    {\mbox{$\frac{y_{_{JB}}}{y_{gen}}$}}
\newcommand{\ydarat}    {\mbox{$\frac{y_{_{DA}}}{y_{gen}}$}}
\newcommand{\yptrat}    {\mbox{$\frac{y_{_{PT}}}{y_{gen}}$}}
\newcommand{\yprrat}    {\mbox{$\frac{y_{(1)}}{y_{gen}}$}}
\newcommand{\yprprrat}  {\mbox{$\frac{y_{(2)}}{y_{gen}}$}}
\newcommand{\yerat}     {\mbox{$\frac{y_{e}}{y_{gen}}$}}
\newcommand{\yptye}      {\mbox{$\frac{y_{_{PT}}}{y_{e}}$}}
\newcommand{\qptrat}    {\mbox{$\frac{Q^2_{PT}}{Q^2_{gen}}$}}
\newcommand{\qdarat}    {\mbox{$\frac{Q^2_{DA}}{Q^2_{gen}}$}}
\newcommand{\qerat}     {\mbox{$\frac{Q^2_{e}}{Q^2_{gen}}$}}
\newcommand{\qprprrat}  {\mbox{$\frac{Q^2_{(2)}}{Q^2_{gen}}$}}
\newcommand{\perc}      {~\mbox{\hspace*{-0.3em}\%}}
\newcommand{\Gev}       {\mbox{${\rm GeV}$}}
\newcommand{\Gevsq}     {\mbox{${\rm GeV}^2$}}
\newcommand{\qsd}       {\mbox{${Q^2}$}}
\newcommand{\x}         {\mbox{${\it x}$}}
\newcommand{\y}         {\mbox{${\it y}$}}
\newcommand{\ye}        {\mbox{${y_{e}}$}}
\newcommand{\smallqsd}  {\mbox{${q^2}$}}
\newcommand{\ra}        {\mbox{$ \rightarrow $}}
\newcommand{\ygen}      {\mbox{${y_{gen}}$}}
\newcommand{\yjb}       {\mbox{${y_{_{JB}}}$}}
\newcommand{\yda}       {\mbox{${y_{_{DA}}}$}}
\newcommand{\qda}       {\mbox{${Q^2_{_{DA}}}$}}
\newcommand{\qjb}       {\mbox{${Q^2_{_{JB}}}$}}
\newcommand{\ypt}       {\mbox{${y_{_{PT}}}$}}
\renewcommand{\xpt}     {\mbox{${x_{_{PT}}}$}}
\newcommand{\xda}     {\mbox{${x_{_{DA}}}$}}
\newcommand{\xp}     {\mbox{${x_p}$}}
\newcommand{\qpt}       {\mbox{${Q^2_{_{PT}}}$}}
\newcommand{\ypr}       {\mbox{${y_{(1)}}$}}
\newcommand{\yprpr}     {\mbox{${y_{(2)}}$}}
\newcommand{\ypps}      {\mbox{${y_{(2)}^2}$}}
\newcommand{\gammah}    {\mbox{$\gamma_{_{H}}$}}
\newcommand{\gammahc}   {\mbox{$\gamma_{_{PT}}$}}
\newcommand{\gap}       {\hspace{0.5cm}}
\newcommand{\gsim}      {\mbox{\raisebox{-0.4ex}{$\;\stackrel{>}{\scriptstyle \sim}\;$}}}
\newcommand{\lsim}      {\mbox{\raisebox{-0.4ex}{$\;\stackrel{<}{\scriptstyle \sim}\;$}}}
\newcommand{\sleq} {\raisebox{-.6ex}{${\textstyle\stackrel{<}{\sim}}$}}
\newcommand{\sgeq} {\raisebox{-.6ex}{${\textstyle\stackrel{>}{\sim}}$}}

\renewcommand{\thefootnote}{\arabic{footnote}}
\date{}
\title{\bf Observation of Scaling Violations\\
in Scaled Momentum Distributions \\
at HERA \\}
\author{ZEUS Collaboration}
\date{}
\maketitle
\vspace{5 cm}
\begin{abstract}
Charged particle production has been measured in deep
inelastic
scattering (DIS) events 
over a large range of $x$ and $Q^2$ using the ZEUS detector.
The evolution of the scaled momentum, $x_p$, with $Q^2,$ in the 
range 10 to 1280 ${\rm GeV}^2$,
has been investigated in the
current fragmentation region of the Breit frame.
The results show clear evidence, in a single experiment, for scaling violations 
in scaled momenta as a function of $Q^2$.

\end{abstract}
\vspace*{-16.5cm}
\leftline{\tt DESY 97-183}
\thispagestyle{empty}
\setcounter{page}{0}
\def\3{\ss}
\pagenumbering{Roman}

\newpage

\begin{center}
{\Large  The ZEUS Collaboration}
\end{center}

  J.~Breitweg,
  M.~Derrick,
  D.~Krakauer,
  S.~Magill,
  D.~Mikunas,
  B.~Musgrave,
  J.~Repond,
  R.~Stanek,
  R.L.~Talaga,
  R.~Yoshida,
  H.~Zhang  \\
 {\it Argonne National Laboratory, Argonne, IL, USA}~$^{p}$
% ...
\newpage
% ...
$^{  23}$ now at ATLAS Collaboration, Univ. of Munich \\
% ...
$^{  32}$ partially supported by the Foundation for German-Russian
Collaboration
DFG-RFBR \\ \hspace*{3.5mm} (grant no. 436 RUS 113/248/3 and no. 436 RUS
113/248/2)\\
% ...
%
\newpage   
%
\begin{tabular}[h]{rp{14cm}}
$^{a}$ &  supported by the Natural Sciences and Engineering Research
          Council of Canada (NSERC)  \\
$^{b}$ &  supported by the FCAR of Qu\'ebec, Canada  \\
$^{c}$ &  supported by the German Federal Ministry for Education and
          Science, Research and Technology (BMBF), under contract
          numbers 057BN19P, 057FR19P, 057HH19P, 057HH29P, 057SI75I \\
$^{d}$ &  supported by the MINERVA Gesellschaft f\"ur Forschung GmbH,
          the German Israeli Foundation, and the U.S.-Israel Binational
          Science Foundation \\
$^{e}$ &  supported by the German Israeli Foundation, and
          by the Israel Science Foundation
  \\
$^{f}$ &  supported by the Italian National Institute for Nuclear Physics
          (INFN) \\
$^{g}$ &  supported by the Japanese Ministry of Education, Science and
          Culture (the Monbusho) and its grants for Scientific Research \\
$^{h}$ &  supported by the Korean Ministry of Education and Korea Science
          and Engineering Foundation  \\
$^{i}$ &  supported by the Netherlands Foundation for Research on
          Matter (FOM) \\
$^{j}$ &  supported by the Polish State Committee for Scientific
          Research, grant No.~115/E-343/SPUB/P03/002/97, 2P03B10512,
          2P03B10612, 2P03B14212, 2P03B10412 \\
$^{k}$ &  supported by the Polish State Committee for Scientific
          Research (grant No. 2P03B08308) and Foundation for
          Polish-German Collaboration  \\
$^{l}$ &  partially supported by the German Federal Ministry for
          Education and Science, Research and Technology (BMBF)  \\
$^{m}$ &  supported by the Fund for Fundamental Research of Russian Ministry
          for Science and Edu\-cation and by the German Federal Ministry for
          Education and Science, Research and Technology (BMBF) \\
$^{n}$ &  supported by the Spanish Ministry of Education
          and Science through funds provided by CICYT \\
$^{o}$ &  supported by the Particle Physics and
          Astronomy Research Council \\
$^{p}$ &  supported by the US Department of Energy \\
$^{q}$ &  supported by the US National Science Foundation \\
\end{tabular}
%

\pagestyle{plain}

\newpage
\pagenumbering{arabic}                   
\setcounter{page}{1}

\section{\label{sec:intro} Introduction}

The observation of scaling violations in 
structure functions~\cite{DISscal}  measured 
in neutral current, deep inelastic
scattering (DIS) helped to 
establish Quantum Chromodynamics (QCD) 
as the theory of strong interactions and has led to 
measurements of the strong coupling constant, $\alpha_s.$
Similar scaling violations are predicted in the fragmentation functions,
which represent the
probability
for a parton to fragment into a particular
hadron carrying a fraction of the parton's energy. 
Fragmentation functions incorporate the long distance, non-perturbative 
physics of the hadronization process in which the observed hadrons are 
formed from final state partons of the hard scattering process
and, like structure functions,
cannot be calculated in perturbative QCD, but
can be evolved from a starting distribution at a defined energy scale.
If the fragmentation functions are combined with 
the cross sections for the inclusive production of 
each parton type in the given physical process, predictions can be made for 
scaling violations in the scaled momentum
spectra of final state hadrons \cite{webnas}.
These scaling violations 
allow a measurement of \alphas\ and such studies have been
performed at LEP~\cite{ALEPH,DELPHI} by incorporating
lower energy PETRA data. 

The event kinematics of DIS
are determined by the negative square of the 4-momentum transfer at
the positron vertex, 
$Q^2\equiv-q^2$, and the Bjorken scaling variable, $x=Q^2/2P \cdot q$, 
where $P$ is the four-momentum of the proton.
In the Quark Parton Model (QPM), 
the interacting quark from the proton carries the four-momentum $xP$.
The variable $y$, the fractional energy transfer to the proton in its rest 
frame, is related to $x$ and $Q^2$ by $y\simeq Q^2/xs$, where $\sqrt s$ is
the positron-proton centre of mass energy.
A natural frame in which to study the dynamics of the hadronic final state
in DIS is the Breit frame~\cite{feyn,walsh}, which has been used in previous
studies of QCD effects~\cite{Zcoh,val,h1old} at HERA.
In this frame the exchanged
virtual boson is purely
space-like with 3-momentum $\vec{q}=(0,0,-Q)$, the incident quark
carries momentum $Q/2$ in the positive $Z$-direction,
and the outgoing struck quark carries Q/2 in the negative
$Z$-direction.  A final state particle has momentum $p^B$ in this frame,
and is assigned to the current region if $p^B_Z$ is negative, and to the
target frame if $p^B_Z$ is positive.
The advantage of this 
frame lies in the maximal separation of the outgoing parton from
radiation associated with the incoming parton and the proton remnant,
thus providing the optimal environment for the study of its fragmentation.

In this analysis the inclusive charged particle
distributions of the scaled momenta, $ x_p,$
 in the current 
region of the Breit frame are measured;
$x_p$ is the momentum
of a track measured in the Breit frame, $p^B$,
scaled by $Q/2$, the maximum possible momentum
(ignoring effects due to the intrinsic $k_T$ of the quark within the
proton). 
 
The scaled momentum distributions are studied as a function of $Q$ and $x$,
in the range
$6\times10^{-4} < x < 5\times10^{-2}$ and \mbox{$10<Q^2<1280~$}  
GeV$^2.$
Thus the evolution of fragmentation functions can be observed within a
single experiment over a wide range of $Q.$
A similar analysis has recently been performed by the H1
collaboration~\cite{h1new}.

\section{\bf Experimental Setup}
\label{s:detector}
 
The data presented here were taken in 1994 at the positron-proton
 collider HERA using the ZEUS detector. During this period HERA operated
with positrons of energy $E_e=27.5$~GeV and protons with energy $820$~GeV.
 The data of this analysis correspond to a 
luminosity of $2.50\pm0.04 {\rm \ pb^{-1}}.$
 A detailed description of the ZEUS detector can be 
found in~\cite{b:sigtot_photoprod,b:Detector};
we present here a brief description of the components most relevant to the
present
 analysis.

Throughout
this paper we use the standard ZEUS right-handed coordinate system, in which
$X = Y = Z = 0$ is the nominal interaction point, the positive
$Z$-axis points in the direction of the proton  beam (referred
to as the forward direction) and the $X$-axis is horizontal, pointing towards
the centre of HERA.
 
The tracking system consists of a vertex detector (VXD)~\cite{b:VXD} 
and a central tracking chamber (CTD)~\cite{b:CTD}
enclosed in a 1.43 T solenoidal magnetic field. 
Immediately surrounding the beampipe is the VXD  which
consists of 120 radial cells, each with 12 sense wires.
The CTD, which encloses the VXD,  is a drift chamber consisting of 
72~cylindrical layers, arranged in 9 superlayers. Superlayers with  
wires parallel to the beam axis alternate with those inclined at a small 
angle to give a stereo view. 
The single hit efficiency of the CTD is greater than 
95$\%$ and the measured resolution in transverse 
momentum for tracks with hits in all the superlayers is 
$ \sigma_{p_{T}}/p_{T} = 0.005 p_{T} \bigoplus 0.016 $ ($p_{T}$ in~GeV).

Outside the solenoid is the uranium-scintillator 
calorimeter (CAL)~\cite{b:CAL}, 
which is divided into three parts: 
forward, barrel and rear covering the polar regions
$2.6^\circ$ to $36.7^\circ$,
$36.7^\circ$ to $129.1^\circ$ and
$129.1^\circ$ to $176.2^\circ$, respectively. 
The CAL covers 99.7$\%$ of the solid angle, with 
holes of $ 20 \times 20 $ cm$^{2}$ in the centres of                       
the forward and rear calorimeters to 
accommodate the HERA beam pipe. Each of the calorimeter parts is subdivided
into towers which are segmented longitudinally into electromagnetic (EMC) 
and hadronic (HAC) sections. These sections are further subdivided into cells
each of which is read out by two photomultipliers. From test beam data, 
energy resolutions 
of ${\sigma}_E/E = 0.18/\sqrt{E}$ for electrons and 
${\sigma}_E/E = 0.35/\sqrt{E}$ for hadrons ($E$ in~GeV) have been obtained.

The small angle rear tracking 
detector (SRTD)~\cite{b:SRTD}, which is attached to 
the front face of the rear calorimeter, measures the impact point of charged
particles at small angles with respect to the positron beam direction.

The luminosity is measured by means of the Bethe-Heitler
process $ep \rightarrow e  \gamma p,$ by detecting the photon in a
calorimeter~\cite{b:LUMI} positioned at $Z = -107 {\rm \ m}$ which has an energy
resolution of $\sigma_E/E$ = 0.18/$\sqrt {E ({\rm GeV})}$
 under test beam conditions.
The luminosity calorimeter is
also used to tag photons from initial state radiation in DIS
events.

\section{Kinematic Reconstruction }
\label{section:CALIB}

The ZEUS detector is almost hermetic, allowing the kinematic
variables $x$ and $Q^2$ to be reconstructed in a variety of
ways using combinations of positron and hadronic system energies and
angles. 
Variables calculated only 
from the measurements of the energy, $E^{\prime}_e,$ 
and angle, $\theta_e,$
of the scattered positrons are denoted with the subscript $e$, whilst those
calculated from the hadronic system measurements, with the Jacquet
Blondel method~\cite{jb}, are denoted by the subscript $JB.$
Variables calculated by these approaches
are only used in the event selection.
In the double angle method~\cite{DA}, denoted by $DA,$  the 
kinematic variables are determined using $\theta_e$
and the angle $ \gamma_H $ (which is the direction of the struck
quark in QPM), defined from the hadronic
final state:

\[\cos\gammah=\frac{(\sum_i p_X)^2 +(\sum_i
p_Y)^2-(\sum_i(E-p_Z))^2}{(\sum_i p_X)^2 +(\sum_i
p_Y)^2+(\sum_i(E-p_Z))^2}\ , \]
where the sums run over all CAL
cells $i$, excluding those assigned to the scattered
positron, and ${\vec p}=(p_X, p_Y, p_Z)$ is the 3-momentum 
assigned to
a cell of energy $E$.  The cell angles are calculated from the
geometric centre of the cell and the vertex position of the event.
This angle is then combined with the measured angle of the scattered
positron to calculate $x$ and $Q^2$. The two angles
can also be used to
recalculate the energy of the scattered positron:

$$ E_{DA}^{\prime} = \frac{2 E_e \sin \gammah }
{\sin \gammah + \sin \theta_e - \sin (\gammah + \theta_e)}.$$

The $DA$ method was used throughout this analysis
for the calculation of the boosts and the kinematic variables
because it is less
sensitive to systematic
uncertainties in the energy measurement than the other methods discussed
above.

An additional method of measuring $y$ and $Q^2$ 
(the PT method~\cite{z_shift}) was used as a systematic check
of the kinematic reconstruction to determine the binning of
the data and the boost vector.
This method uses a more sophisticated
 combination of the information from the measurements of both the
hadronic system and the positron.

In order to boost to the Breit frame it is necessary to calculate 
the velocity of the Breit frame with respect to
the laboratory frame, which is given by
$ \vec{\beta} =  (\vec{q} + 2 x \vec{P})/( q_{0} + 2 x P_{0})  $
where ($q_{0},\vec{q}$) and ($P_0,\vec{P}$) are
the 4-momenta of the exchanged photon and the incident proton beam
respectively.
The event is then rotated so that the virtual photon is along the
negative
$Z$-axis.
The Breit frame boost vector
was reconstructed using 
$E_{DA}^{\prime}$ and the 
polar and azimuthal angles measured from the impact point of the
scattered $e^+$ on the calorimeter or SRTD.
The 4-momentum vectors of the charged particles were boosted to the Breit 
frame, assuming the pion mass to determine the particle's energy, and 
were assigned to the current region if $p^B_Z<0.$

\section{Event Reconstruction  and Selection}
\label{sec:recons}

The triggering and online
event selections were identical to those used for the 
measurement of the structure function $F_2$~\cite{z_shift}.

To reduce the effects of noise due to the uranium radioactivity
on the calorimetric measurements
all EMC (HAC) cells with an energy deposit of less than 60 (110)~MeV 
were discarded from the
analysis. For cells with isolated energy deposits
 this cut was increased to 100 (150)~MeV. 

The track finding algorithm starts with
hits in the  outer axial superlayers of the CTD.  As the
trajectory is followed inwards towards the beam axis,
hits  from inner superlayers are added to the track. 
The momentum vector is determined in a 5-parameter helix fit.
The reconstructed tracks used in this analysis
are associated with the primary event vertex 
and have $p_T>150$~MeV and $|\eta|<1.75,$
where $\eta$ is the pseudorapidity given by $-\ln(\tan(\theta/2))$ with
$\theta$  being the polar angle of the measured track with respect to
the proton direction.
This is a region of high CTD acceptance 
where the detector response and systematics are best understood.

Further selection
criteria were applied both
to ensure accurate reconstruction of the kinematic
variables and to increase the purity of the sample by
eliminating background from photoproduction processes. These cuts were:
\begin{itemize}
\item
$E^\prime_e \ge 10~{\rm GeV}$,
to achieve a high purity sample of DIS events;
\item
$Q^2_{DA}\geq 10$ GeV$^2$;
\item
$y_e\leq 0.95$,
to reduce the photoproduction background;
\item
$y_{JB}\geq 0.04$, to give sufficient accuracy for $DA$ reconstruction;
\item $35 \le \delta = \sum_i\left( E-p_Z\right)\le 60$~GeV summed over
all calorimeter cells, to remove photoproduction events and
      events with large radiative corrections.
\item $ |X|  >  16\ {\rm cm}\ {\rm or}\ |Y|  >  16\ {\rm cm} ,$ where 
  $X$ and $Y$ are the impact position of the 
  positron on the CAL as determined using the SRTD 
  to avoid
  the region directly adjacent to the rear beam pipe.
\item $ -40  <  Z_{\rm vertex}  <  50\ {\rm cm},$ to reduce background
events from non $ep$ collisions.
\end{itemize}

In total, 68066 events satisfy the above cuts and 
are reconstructed
in the $(x,Q^2)$ bins (as calculated by the $DA$ method) 
that are listed in Table~\ref{tab:bins}.
 The sizes of the bins in $x$ and $Q^2$ 
 were chosen to have good statistics in each
 bin and to limit the migrations between bins~\cite{Zcoh}.

\section{QCD Models and Event Simulation}
\label{s:model}
Monte Carlo event simulation is used to correct for acceptance and
resolution effects.  The detector simulation is based on the
GEANT~3.13~\cite{GEANT} program and incorporates our best knowledge of the
apparatus. 

To calculate the acceptance, neutral current DIS events 
were generated, via the DJANGO \mbox{program}~\cite{DJANGO},
using HERACLES \cite{HERACLES} which 
incorporates first order electroweak corrections. 
The QCD cascade was modelled 
with the colour-dipole model including the boson-gluon fusion process,
using the ARIADNE~4.08 \cite{ariadne} program. In this model coherence
effects are implicitly included in the formalism of the parton cascade.
This program uses the Lund string fragmentation model \cite{string} 
for the hadronisation phase, as implemented 
in JETSET~7.3 \cite{JETSET}. 
Two Monte Carlo samples were generated, $1.6~{\rm pb}^{-1}$ 
with $Q^2>3$ GeV$^2$ and 
$2.6~{\rm pb}^{-1}$ 
with $Q^2>6$ GeV$^2,$ using the 
$\rm{MRSA}$ \cite{mrsa} parameterisation of the parton distribution functions.
Another approach to modelling the parton cascade is included in
the LEPTO(6.5\_1)~\cite{LEPTO}
program, which incorporates the LO matrix element
matched to parton showers
(MEPS). This program also uses the Lund string fragmentation model
and was used for the generator level calculations used
in comparisons to our data.

For the studies of the systematics,
two additional samples of events were 
generated ($1.2~{\rm pb}^{-1}$ with $Q^2>4$
GeV$^2$ and $3.2~{\rm pb}^{-1}$ with  $Q^2>100$ GeV$^2$) 
using the HERWIG~5.8c Monte Carlo
\cite{herwig}, 
where no electroweak radiative 
corrections were applied. 
In HERWIG, coherence effects in the QCD cascades are included by
angular
ordering of successive parton emissions 
and a clustering model is used for the hadronisation \cite{webber,cluster}. 
For this sample the parameterisation of the parton distribution
functions was the $\rm{MRSA^{\prime}}$ set~\cite{mrsap}.
The $\rm{MRSA}$ and $\rm{MRSA^{\prime}}$ parameterisations  
have both been shown to describe reasonably well the HERA measurements
of the proton structure function $F_2$ in the $(x,Q^2)$
range of this analysis~\cite{f2,h1f2}. 


\section{Correction Procedure}

Monte Carlo event generator studies were used to determine 
the mean charged particle acceptance 
in the current region as a function of $(x,Q^2)$.
The chosen analysis intervals in $(x,Q^2)$ correspond to regions of
high acceptance, $74$ to $96\%,$ 
in the current region of the Breit frame.

\begin{table}[hbt]
\begin{center}
\begin{tabular}{|c|c|c|}
\hline
  $x_{DA}$ range &  $Q^2_{DA}$ (GeV$^2$) range  &  No. of events \\
\hline\hline
$6.0 - 12.0 \times 10^{-4} $ & $ 10 - 20 $  & 13898 \\
\hline
$1.2 - 2.4 \times 10^{-3} $ & $ 10 - 20 $  & 11899 \\
   & $ 20 - 40 $   & 8484 \\
   & $ 40 - 80 $   & 5093 \\
\hline
$2.4 - 10.0 \times 10^{-3} $ & $ 20 - 40 $   & 9399 \\
   & $ 40 - 80 $  & 9493 \\
   & $ 80 - 160 $  & 5031 \\
   & $ 160 - 320 $ & 1369 \\
\hline
$1.0 - 5.0 \times 10^{-2} $ & $ 160 - 320 $   & 2131 \\
    & $ 320 - 640 $   & 916 \\
   & $ 640 - 1280 $   & 353 \\
\hline\hline
\end{tabular}
\caption{{\small The ($x$,$Q^2$) analysis bins showing the 
the numbers of accepted events in the ($x$,$Q^2$) bins as 
reconstructed by the $DA$ method.  }}
\label{tab:bins}
\end{center}
\end{table}

Uncertainty in the reconstruction of the boost vector, $\vec{\beta}$, 
was found to be the most significant factor 
on the resolution of $x_p$ and it
leads to the choice of variable bin width in $x_p.$ 
Migration of tracks from the current region to the target region 
was typically $\simeq 6{\rm\ to\ }8\%$ of the tracks generated in the current
region. 
Migrations into the current region from
the target fragmentation region are typically less than 5\% of the 
tracks assigned
to the current region for $ Q^2 > 320 {\rm\ GeV^2}.$
For $10 < Q^2 < 320 {\rm\ GeV^2}$
these migrations are generally of the order of 10 to 15\% .
At $Q^2 < 40  {\rm\ GeV^2}$ and
low values of $y$ this assignment can be as high as
$\simeq 25\%,$  since in the low $y$ region the hadronic activity is low 
and the measurement of $\gamma_H$ becomes distorted by noise in the 
calorimeter leading to a worse $x$ resolution 
and hence an uncertainty in $\vec{\beta}$. 

The correction procedure is based on the detailed Monte Carlo
simulation of the ZEUS detector with the event generators
described in the previous section.
Since the ARIADNE model gives the best overall description of our observed 
energy flow~\cite{zeus:efl} it is used for the standard corrections 
to the distributions.

The data are corrected for trigger and event selection cuts; 
event migration between ($x,Q^2$) intervals; 
QED radiative effects; 
track reconstruction efficiency; 
track selection cuts in $p_T$ and $\eta$; 
track migration between the current and target regions; 
and for the 
products of $\Lambda$ and $K^{0}_S$ decays which are assigned to the 
primary vertex. 

Correction factors were obtained from the Monte Carlo simulation by 
comparing the generated distributions, without $\Lambda$ and $K^0_S$
decay products, with the reconstructed 
distributions after the detector and trigger
simulations followed 
by the same reconstruction, selection and analysis
as the real data.
The 
correction factors, $F(\xp\!)$, were calculated for each $x_p$~bin with
the formula:
$$
F(\xp\!) = \frac{1}{N_{\rm gen}} \; \left( \frac{dn}{d\xp} \right) _{\rm gen} 
 \left/ \frac{1}{N_{\rm obs}} \; \left( \frac{dn}{d\xp} \right) _{\rm obs}  \right.
$$
where $N_{\rm gen}$ ($N_{\rm obs}$) is the number of generated (observed) 
Monte Carlo events in each $(x,Q^2)$ interval and $n$ is the number of
charged particles (tracks) in the current region in the corresponding
$(x,Q^2)$ interval.
The  correction factors
are typically between 1.0 and 1.3, except 
in the lowest $x_p$ bin $( 0.02 < x_p < 0.05)$ of the two lowest $Q^2$
bins where they are about $1.5.$

\section{Results}

\begin{figure}[hbt]
\begin{center}\mbox{\epsfig{file=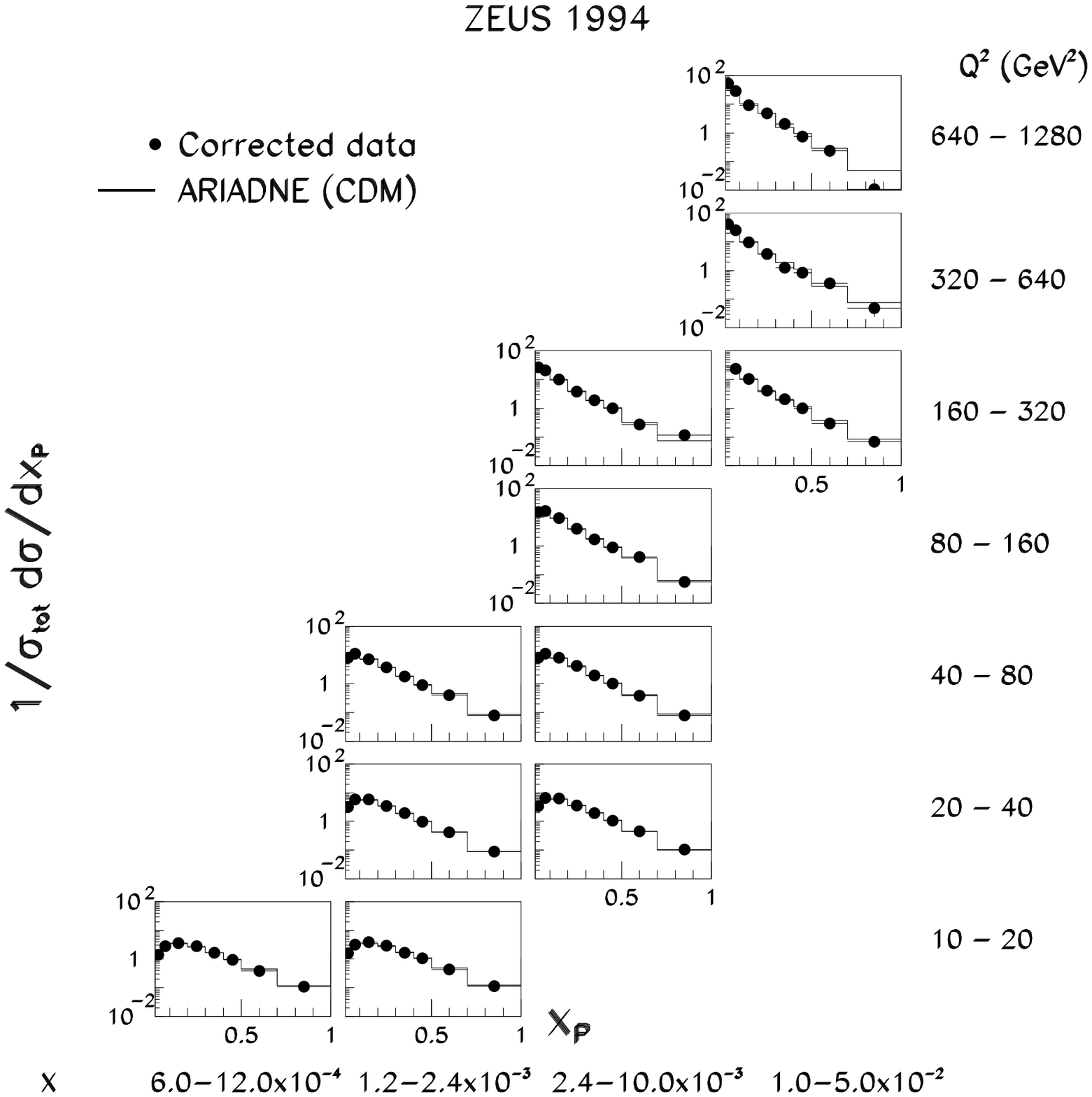,width=17cm}}
\end{center}
\caption{\small The inclusive charged particle
distributions, $ 1/\sigma_{tot}~ d\sigma/dx_p$, 
in the current fragmentation region of the Breit frame
compared with the generated distributions 
from the Monte Carlo (ARIADNE 4.08).
The statistical errors are generally within the size of the points.} 
\label{corrdat}
\end{figure}

The inclusive charged particle distributions, $1/\sigma_{tot}
d\sigma/d x_p$ where $\sigma_{tot}$ is the DIS cross section in the
chosen $(x,Q^2)$ bin, 
are shown in figure~\ref{corrdat}.
In the low $(x,Q^2)$ regions these distributions
peak at $x_p \approx 0.2$ with a broad tail
out to $1.$ As $Q^2$ increases 
the data fall off more sharply from the minimum $x_p.$
In the same figure the ARIADNE Monte Carlo generator predictions are
also shown, which are in good agreement with the data.
The increasingly steep fall-off towards
higher values of $x_p$ as $Q^2$ increases
corresponds to  the production of
more particles with 
a smaller fractional momentum, and
 is indicative of scaling
violation in the fragmentation function. 
These scaling violations
 can be seen more clearly if the data are plotted in bins of
fixed $x_p$ as a function of $Q^2.$
Figure~\ref{syst_hrw} shows the charged 
particle distributions with statistical errors 
combined in quadrature with the systematic errors which will be
discussed below.
For $Q^2 > 80{\rm\ GeV^2}$ the distributions rise
with $Q^2$ at low $x_p$ and 
fall-off at high $x_p$ and high $Q^2$.
By measuring the amount of
scaling violation one can ultimately
measure the amount of parton radiation and
thus determine $\alpha_s.$
Below $Q^2=80{\rm\ GeV^2}$ the fall-off is due to depopulation of the
current region discussed later.

\subsection{Systematic Studies}

The systematic uncertainties
 in the measurement can be divided into three classes: 
errors due to event reconstruction and selection; to track reconstruction and
selection; and to the Monte Carlo generator used.
\begin{itemize}
\item {\bf Event Reconstruction and Selection}\\
\noindent The event selection procedure was checked by altering the $y_e, y_{JB}$ and
$\delta$ cuts: the resulting
shifts of the corrected distributions were small,
with all points moving systematically in the same direction at the 
$\pm 1{\rm\ to\ }3 \%$ level.
By removing the noise suppression, described in section~\ref{sec:recons},
a systematic shift at the $\pm 1{\rm\ to\ }3 \%$ level is produced.
Using the boost calculated from the PT method
produced a larger shift:
for $Q^2 < 160 {\rm \ GeV^2}$ the
systematic shift is $+10{\rm\ to\ }15\%$ at large $x_p,$ whilst
$+2{\rm\ to\ }5\%$ at small $x_p;$
for $Q^2 \ge 160 {\rm \ GeV^2}$ the systematic
shift is constant with
$x_p$ at the  $+2{\rm\ to\ }5\%$ level. 
\item {\bf Track Reconstruction and Selection}\\
\noindent The major systematic on 
the track reconstruction was obtained considering all reconstructed
tracks as opposed to only those assigned to the primary vertex.
This produced a typical shift of
2{\rm\ to\ }6\% with no
systematic dependence on the value of $x_p,$ except in the bin 
$0.02 < x_p < 0.05$ at low $Q^2$ and $x$ where the systematic shift was up to 
$+15\%.$
Tightening the tracking cuts on $|\eta|$ and $p_T$
particularly affected those $(x,Q^2)$ bins 
where the track acceptance is strongly 
dependent on the negative $\eta$ cut.
The general trend is for the cross section to be 
shifted to higher values. The bins most affected have an average shift of 
up to $5\%$, whilst in general the shift is $\sleq 1\%.$
\item {\bf Monte Carlo Generator}\\
\noindent Using a different Monte Carlo generator
(HERWIG rather than ARIADNE) 
led to distributions which were systematically lower by about $10 \%$ in 
the range of $0 \leq x_p \leq 0.3$. 
In the range $ x_p > 0.5$, 
the HERWIG corrected 
distributions were systematically higher (lower) by 5{\rm\ to\ }10 \% 
for $Q^2 < 40~\mbox{GeV}^2$ ($Q^2 > 100~\mbox{GeV}^2 $).
In the range $0.3 \leq x_p \leq 0.4$ the two generators gave results that
were in good agreement.
\end{itemize}

All positive (negative)
systematic shifts in each of the $x_p$ bins were combined in quadrature to
give an estimate of the overall positive (negative) 
systematic uncertainty on the measurement.
These systematic shifts do not affect the observation
of scaling violations.

\subsection{Discussion}
\begin{figure}[htb]
\begin{center}\mbox{\epsfig{file=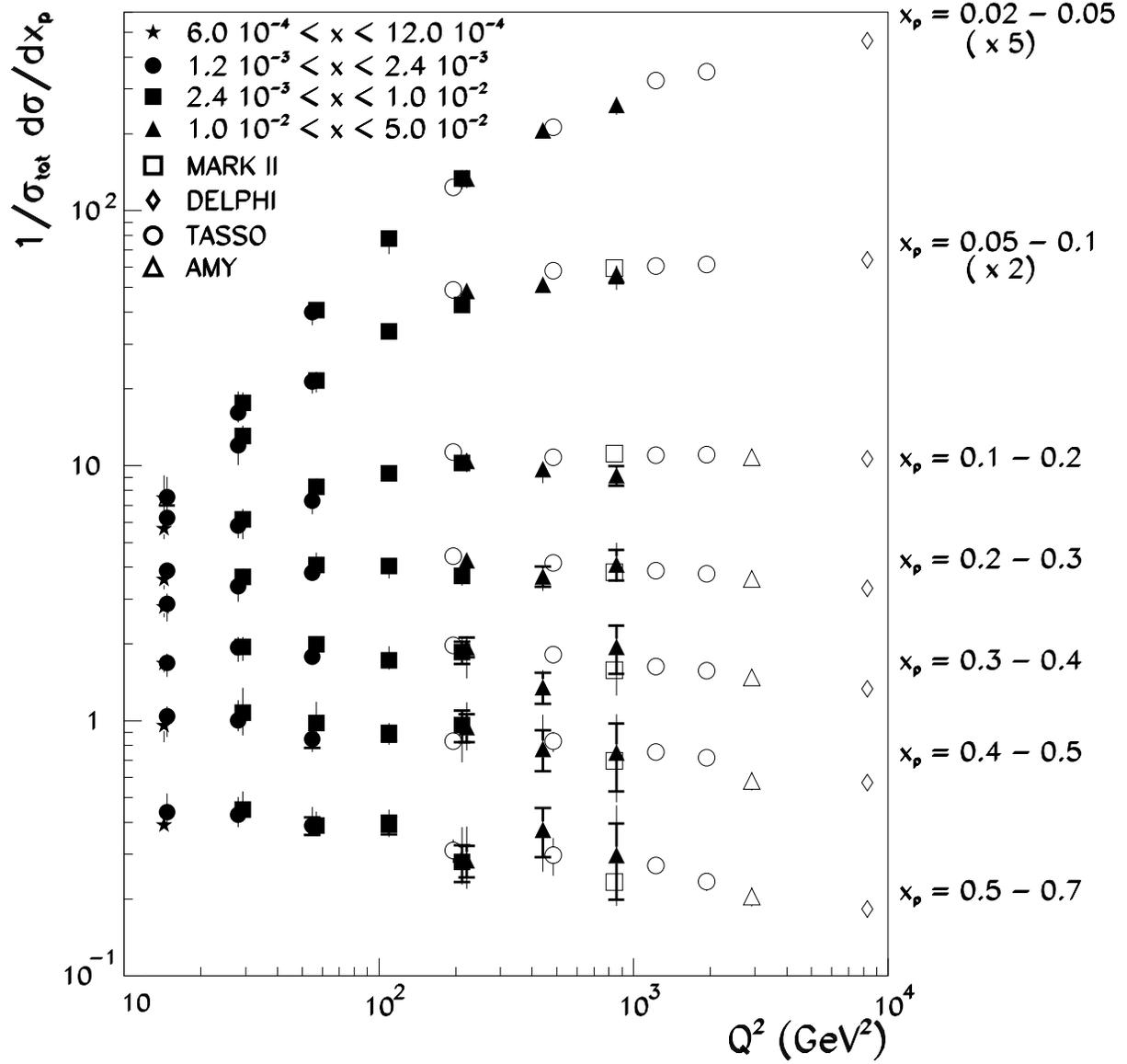,width=17cm}}
\end{center}
\caption{\small The inclusive charged particle distribution, 
$ 1/\sigma_{tot}~ d\sigma/dx_p$, 
in the current fragmentation region of the Breit frame. 
The filled data points are from ZEUS.
The full errors are statistical and systematic combined in quadrature.
The thick lines denote the statistical error. 
The open points represent data from $e^+e^-$ experiments divided by two to
account for $q$ and $\bar q$ production (also corrected
for contributions from $K^0_S$ and $\Lambda.$)}
\label{syst_hrw}
\end{figure}

Fragmentation in DIS of a quark carrying momentum $Q/2$ in the Breit
frame may be compared to fragmentation in $e^+e^-$ annihilation of
the produced $q$ and $\bar q,$ each carrying momentum $\sqrt{s_{e^+e^-}}/2.$
In  figure~\ref{syst_hrw} the ZEUS data are compared at $Q^2=s_{e^+e^-}$
to $e^+e^-$ data~\cite{eedata},
divided by two to account for the production of both a $q$ and $\bar q.$
The $e^+e^-$ data have also been corrected by $\simeq 8\%$
for the decay products of
$\Lambda$ and $K^0_S,$ using the JETSET~7.3 Monte Carlo tuned to DELPHI
data.
In the $Q^2$ range shown there is good agreement between the
current region of the Breit frame in DIS and the $e^+e^-$ experiments.
When our data are compared with the lower energy
 SPEAR~\cite{Patrick} data ($\sqrt{s}=5.2,6.5\ {\rm GeV,}$ not shown)
discrepancies begin to show up. 
They can be explained in terms of the kinematic depopulation
of the current region described in the next paragraph.

The turnover observed in the ZEUS data
at low $x_p$ and low $Q^2,$ figure~\ref{syst_hrw}, can be attributed to 
processes not present in $e^+e^-$ (eg scattering off a sea quark and/or
boson gluon fusion)
depopulating the current region~\cite{val}. 
This is best illustrated when discussing the production of a pair
of partons in DIS with a large invariant mass, $\hat{s}$~\cite{walsh}.
When $Q^2 \gg \hat{s}$, the radiation is emitted in the direction of
the struck quark in the QPM.
However, at low $Q^2$ and low
$x$, $\hat{s}$ is likely to be bigger than $Q^2$ and the radiation will be
emitted in the direction opposite to the QPM direction and will
therefore
not be in the current region as defined in section~\ref{sec:intro}. 
In particular, the boson-gluon fusion diagram,
which dominates at low-$x,$ provides a significant cross section for large
mass radiation~\cite{2jets}, thus producing this depopulation.

\begin{figure}[htb]
\begin{center}\mbox{\epsfig{file=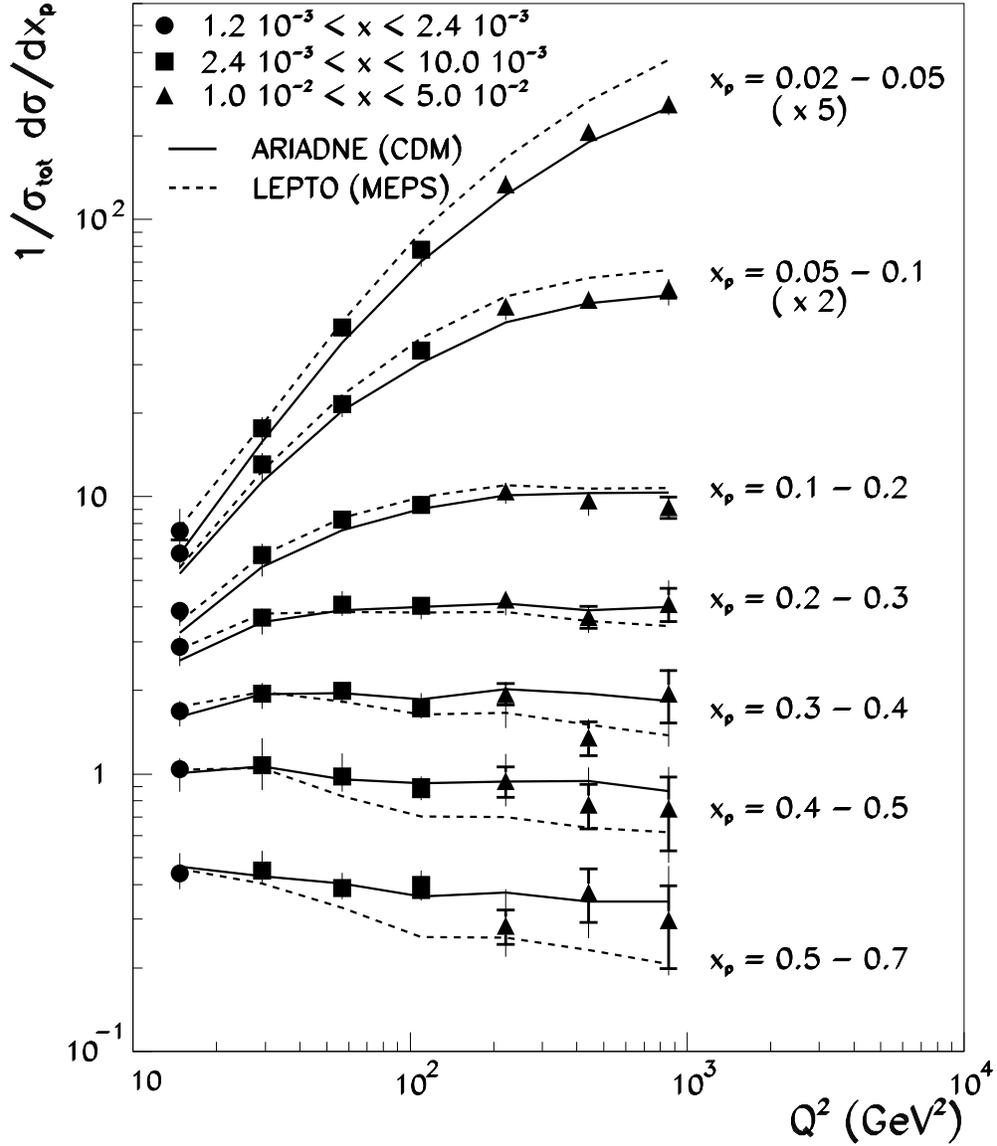,width=17cm}}
\end{center}
\caption{\small The inclusive charged particle distribution, 
$ 1/\sigma_{tot}~ d\sigma/dx_p$, 
in the current fragmentation region of the Breit frame. 
The data points are from ZEUS.
The full errors are statistical and systematic combined in quadrature.
The thick lines denote the statistical error. The curves represent
leading-log Monte Carlo models: the full line is ARIADNE (CDM) and the dashed
line is LEPTO (MEPS).
For clarity of presentation, only the higher $x$ bin is shown in
each $Q^2$ interval.}
\label{data_mc}
\end{figure}

In figure~\ref{data_mc} the data are compared to
two leading-log Monte Carlo predictions which are
implemented within the Lund fragmentation
framework. 
Our data are well described by 
the ARIADNE generator, colour dipole model (CDM), 
over the full range of $Q^2.$  This is not true for 
the LEPTO, matrix element+parton showers (MEPS),
model. In particular LEPTO 
overestimate the data at low $x_p$ and there is a greater $Q^2$
dependence than that suggested by the data.
A feature of
both Monte Carlo models is a trend that, at fixed $Q^2$, the
 charged particle distribution increases with $x.$ Such a trend is
also observed in the data, see fig~\ref{syst_hrw}.
No tuning of the Monte Carlo parameters has been performed
to improve the agreement with the data.

\begin{figure}[htb]
\begin{center}\mbox{\epsfig{file=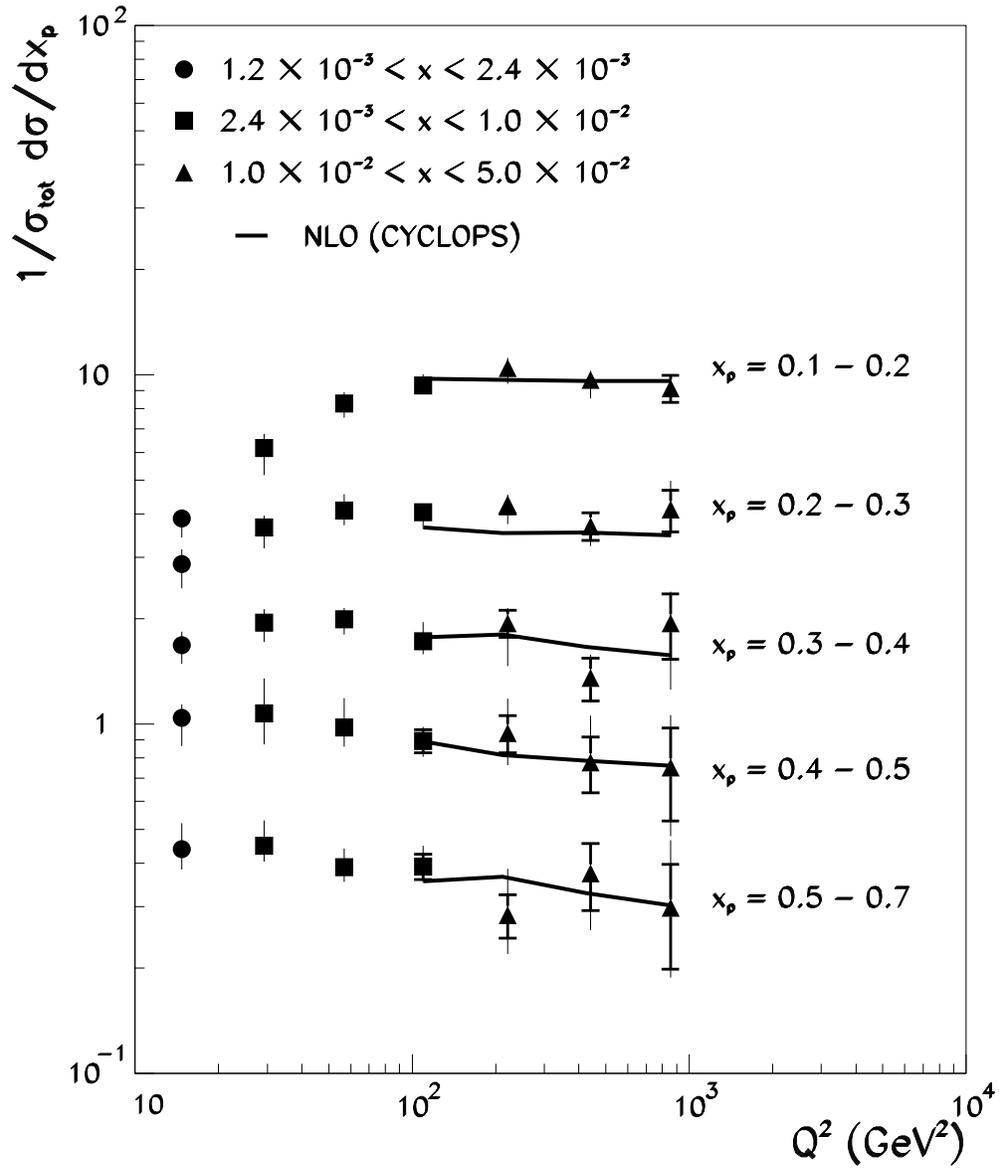,width=17cm}}
\end{center}
\caption{\small The inclusive charged particle distribution, 
$ 1/\sigma_{tot}~ d\sigma/dx_p$, 
in the current fragmentation region of the Breit frame compared to the
NLO calculation, CYCLOPS~\protect\cite{Dirk}.
For clarity of presentation, only the higher $x$ bin is shown in
each $Q^2$ interval.}
\label{NLO}
\end{figure}

Our results can be compared to 
the next-to-leading order (NLO)
QCD calculations, as implemented in CYCLOPS~\cite{Dirk}, of the charged
particle inclusive distributions
in the restricted region
$Q^2 > 80 {\rm \ GeV^2}$  and
$ x_p > 0.1 ,$
where the theoretical uncertainties are small and unaffected by the 
hadron mass effects which are not included in the fragmentation 
function~\cite{Dirk1}. 
This comparison is shown in figure~\ref{NLO}.
The NLO calculation combines a full next-to-leading order matrix element
with the
${\rm MRSA^{\prime}}$ parton densities (with $\Lambda_{\rm QCD} =
230{\rm \ MeV})$
and NLO fragmentation functions
derived by
Binnewies et al. from fits to $e^+e^-$ data \cite{binnewies}.
The  data and the NLO calculations are in good agreement, 
supporting the idea of universality of quark fragmentation. 


\section{Conclusions}
Charged particle distributions have been studied
in the current region of the Breit frame in DIS over a wide range
of $Q^2$ values.
These results show clear evidence in a single experiment
for scaling violations 
in scaled momenta 
as a function of $Q^2$.
The data are also well described with NLO calculations. 

The comparison between  $e^+e^-$ data at $Q^2 = s_{e^+e^-}$
and the current region of the Breit frame in DIS for $Q^2 > 80{\rm \ GeV^2}$ 
shows also a good agreement.
The observed charged particle spectra are consistent 
with the universality of quark fragmentation in $e^+e^-$ and DIS.

\vskip 2.cm \leftline{\Large\bf Acknowledgements} \vskip4.mm \noindent
We appreciate the contributions to the construction and maintenance of the
ZEUS detector by many people who are not listed as authors.
We thank the DESY computing staff for providing the data analysis environment.
The HERA machine group is especially acknowledged for
the outstanding operation of the collider.
The strong support and encouragement of the DESY Directorate                   
has been invaluable. 

We would like to thank D. Graudenz for valuable discussions and
for providing us with the NLO program, CYCLOPS.

\setcounter{secnumdepth}{0} 


\newpage

\newpage

\newpage

\newpage
\end{document}